\documentclass{elsart}

\usepackage{graphicx}

\usepackage{amssymb}

\usepackage[english]{babel}
\usepackage{scrpage,eurosym,amsmath,latexsym}
\usepackage{graphicx}
\usepackage[latin1]{inputenc}
\usepackage{color}
\begin{document}

\begin{frontmatter}



\title{Scaling Laws in the Spatial Structure \\ of Urban Road Networks}


\author[cor1]{Stefan L\"{a}mmer}
\author[cor1]{, Bj\"{o}rn Gehlsen}
\author[cor1,cor2]{, and Dirk Helbing}


\address[cor1]{Technische Universit\"{a}t Dresden, \\A.-Schubert-Str. 23, D-01062 Dresden, Germany}
\address[cor2]{Collegium Budapest -- Institute for Advanced Study, \\Szenth\'aroms\'ag u. 2, H-1014 Budapest, Hungary}

\begin{abstract}
The urban road networks of the 20 largest German cities have been
analysed, based on a detailed database providing the geographical
positions as well as the travel-times for network sizes up to 37,000
nodes and 87,000 links. As the human driver recognises travel-times
rather than distances, faster roads appear to be `shorter' than
slower ones. The resulting metric space has an effective dimension
$d>2$, which is a significant measure of the heterogeneity of road
speeds. We found that traffic strongly concentrates on only a small
fraction of the roads. The distribution of vehicular flows over the
roads obeys a power-law, indicating a clear hierarchical order of
the roads. Studying the cellular structure of the areas enclosed by
the roads, the distribution of cell sizes is scale invariant as
well.
\end{abstract}


\end{frontmatter}

\vspace{20mm} \emph{If you want to cite this report, please use the
following
reference instead:} \\
S. L\"ammer, B. Gehlsen, and D. Helbing (2006) Scaling laws in the
spatial structure of urban road networks, \emph{Physica A} {\bf
363}(1) pp. 89-95 \vspace{20mm} \pagebreak
\section{Introduction}
\label{sec:Intro}

The scientific interest in network analysis has been steadily
growing since the revolutionary discoveries of Watts and Strogatz
\cite{WattsStrogatz98} and Barab\'asi and Albert \cite{Barabasi99}.
They found out that many real-world networks such as the internet
and social networks exhibit a scale-free structure characterised by
a high clustering coefficient and small average path lengths. The
path lengths, however, are usually not related to geographical
distances. Surprisingly, little attention has been paid to the
spatial structure of networks, even though distances are very
crucial for logistic, geographical and transportation networks.

Urban road networks with links and nodes representing road segments
and junctions, respectively, exhibit unique features different from
other classes of networks
\cite{Buhl05,Crucitti05,Gastner04,Jiang04,Newman02,PortaPrimal}. As
they are almost planar, they show a very limited range of node
degrees. Thus, they can never be scale-free like airline networks or
the internet \cite{Gastner04}. Nevertheless, there exists an
interesting connection between these scale-free networks on the one
hand and road networks on the other hand, since both are extreme
cases of an optimisation process minimising average travel costs
along all shortest paths, given a set of nodes and a total link
length. The properties of the resulting networks strongly depend on
the links' cost function. If the travel costs on all links were
equal, small-world networks with a hub-and-spoke architecture
typical for airline networks or the internet would emerge. However,
with travel costs proportional to the link length, the resulting
networks would exhibit properties typical for road networks
\cite{Gastner04}.

\begin{figure}[ht]
    \begin{center}
        \includegraphics[height=7.0cm]{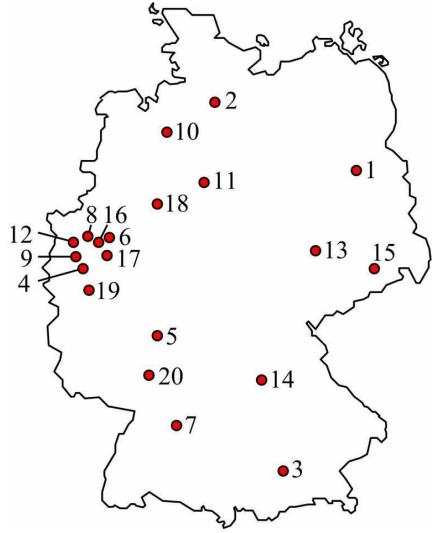}
    \end{center}
    \caption{
        Germany with its 20 largest cities, ranked by their population.
    }
    \label{fig:germany}
\end{figure}

We have extracted road network data of the administrative areas of
the 20 largest German cities from the geographical database Tele
Atlas MultiNet\texttrademark \cite{Teleatlas}, typically used for
real-time navigation systems or urban planning and management. The
data provide a geo-coded polygon for each road segment as well as a
series of properties, e.g. the length, the average expected
travel-time, the speed limit, the driving directions etc. Junctions
and homogeneous road segments of the selected areas are represented
by nodes and links of a corresponding directed graph. The location
of the cities within Germany and the corresponding networks sizes
are shown in Fig.~\ref{fig:germany} and Table~\ref{tab:cities}.
Since the road network of Hanover, ranked $11^\mathrm{th}$, could
not be extracted unambiguously, it was excluded from our analysis.

\newpage

\hspace{1mm}

\begin{table}[hb]
    \renewcommand{\baselinestretch}{0.85}\normalsize
    \centering \scriptsize
    \begin{tabular}{rlcccccccccc}
        \hline
            No.
        &   City
        & \rotatebox{90}{\hbox{\parbox{20mm}{
            Population
            \vspace{1pt}\\
            in 1000
        }}}
        & \rotatebox{90}{\hbox{\parbox{20mm}{
            Area
            \vspace{1pt}\\
            in $\mathrm{km^2}$
        }}}
        & \rotatebox{90}{\hbox{\parbox{20mm}{
            No. of nodes
        }}}
        & \rotatebox{90}{\hbox{\parbox{20mm}{
            No. of links
        }}}
        & \rotatebox{90}{\hbox{\parbox{20mm}{
            Fraction of
            \vspace{1pt}\\
            trees in \%
        }}}
        & \rotatebox{90}{\hbox{\parbox{20mm}{
            Effective
            \vspace{1pt}\\
            dimension $\delta$
        }}}
        & \rotatebox{90}{\hbox{\parbox{20mm}{
            Betweenness
            \vspace{1pt}\\
            exponent $\beta$
        }}}
        & \rotatebox{90}{\hbox{\parbox{20mm}{
            Gini index $g$
        }}}
        & \rotatebox{90}{\hbox{\parbox{20mm}{
            Cell size
            \vspace{1pt}\\
            exponent $\alpha$
        }}}
        & \rotatebox{90}{\hbox{\parbox{20mm}{
            Form factor
            \vspace{1pt}\\
            variance $s_{\phi}$
        }}}
        \\  \hline
        1  &Berlin     &  3,392& 891&   37,020&   87,795&   11.55 & 2.330& 1.481& 0.871& 2.158& 0.159\\
        2  &Hamburg    &  1,729& 753&   19,717&   43,819&   11.93 & 2.350& 1.469& 0.869& 1.890& 0.164\\
        3  &Munich     &  1,235& 311&   21,393&   49,521&   10.74 & 2.463& 1.486& 0.869& 2.114& 0.159\\
        4  &Cologne    & ~\,969& 405&   14,553&   29,359&   21.27 & 2.372& 1.384& 0.875& 1.922& 0.165\\
        5  &Frankfurt  & ~\,644& 249& ~\,9,728&   18,104&   16.90 & 2.388& 1.406& 0.873& 2.009& 0.169\\
        6  &Dortmund   & ~\,591& 281&   10,326&   22,579&   22.82 & 2.091& 1.340& 0.875& 1.809& 0.166\\
        7  &Stuttgart  & ~\,588& 208&   10,302&   21,934&   23.30 & 2.008& 1.377& 0.894& 1.901& 0.170\\
        8  &Essen      & ~\,585& 210&   11,387&   24,537&   22.80 & 2.243& 1.368& 0.892& 1.932& 0.169\\
        9  &Düsseldorf & ~\,572& 218& ~\,8,237&   16,773&   19.35 & 2.700& 1.380& 0.849& 1.964& 0.175\\
        10 &Bremen     & ~\,543& 318&   10,227&   21,702&   23.98 & 2.220& 1.351& 0.909& 1.931& 0.166\\
        11 &Hanover    & ~\,517& 204& ~\,1,589& ~\,3,463&    ---  &  --- &  --- &  --- &  --- &   ---\\
        12 &Duisburg   & ~\,509& 233& ~\,6,300&   14,333&   17.57 & 2.050& 1.480& 0.900& 1.924& 0.169\\
        13 &Leipzig    & ~\,495& 293& ~\,9,071&   21,199& ~\,6.78 & 2.304& 1.320& 0.880& 1.926& 0.153\\
        14 &Nuremberg  & ~\,493& 187& ~\,8,768&   18,639&   19.68 & 2.399& 1.420& 0.854& 1.831& 0.172\\
        15 &Dresden    & ~\,480& 328& ~\,9,643&   22,307&   20.45 & 2.205& 1.355& 0.870& 1.892& 0.156\\
        16 &Bochum     & ~\,389& 146& ~\,6,970&   15,091&   22.19 & 2.279& 1.337& 0.847& 1.829& 0.171\\
        17 &Wuppertal  & ~\,364& 168& ~\,5,681&   11,847&   27.75 & 2.040& 1.279& 0.881& 1.883& 0.162\\
        18 &Bielefeld  & ~\,325& 259& ~\,8,259&   18,280&   26.44 & 2.337& 1.337& 0.872& 1.735& 0.161\\
        19 &Bonn       & ~\,309& 141& ~\,6,365&   13,746&   25.73 & 2.134& 1.374& 0.889& 2.018& 0.173\\
        20 &Mannheim   & ~\,309& 145& ~\,5,819&   12,581&   17.79 & 2.114& 1.455& 0.897& 1.959& 0.162\\
        \hline\\
    \end{tabular}
    \renewcommand{\baselinestretch}{1.00}\normalsize
    \caption{The 20 largest cities of Germany
    and their characteristic coefficients referred to in the following sections.}
    \label{tab:cities}
\end{table}

\newpage

\section{Effective Dimension}
\label{sec:dimension}

In transportation networks with strong geographical constraints, it
is observed that the sizes of neighbourhoods grow according to a
power-law \cite{CsanyiSzendroi04}. We study properties and
implications of such scaling in urban road networks, where distances
are, with respect to human driver's recognition, related to travel
times. Human travel behaviour underlies the universal law of a
constant energy budget \cite{Koelbl03}. The cost of travel must,
therefore, not be measured in the number of road meters, but in the
amount of energy or, assuming a single mode of transport with a
constant energy consumption rate, e.g. car driving, in units of
travel-time. Interestingly, this implies that routes along faster
roads appear `shorter' than slower ones in terms of travel-time. A
distant but well accessible destination is virtually closer than a
near one with a longer time to access. The virtual compression of
faster and the dilation of slower roads result in an effective
deformation of the urban space, whose metric structure we're going
to study.

For any node in the road network, the number of nodes reachable with
a given travel-\emph{distance} budget $r$, i.e. the number of nodes
to which the shortest path is shorter than $r$, essentially scales
with a maximum exponent of 2. This fact is independent of whether
the graph is planar in a strict sense or the urban landscape is
uneven. Considering shortest paths with respect to
travel-\emph{time} instead, the number of nodes $N_v(\tau)$
reachable with a travel-\emph{time} budget $\tau$ follows a scaling
law $N_v(\tau) \sim \tau^\delta$ with $\delta$ being significantly
larger than 2 for all road networks under consideration, see Table
\ref{tab:cities}. The scaling exponent $\delta$ is called the
effective dimension \cite{Gastner04}. The existence of arterial
roads with road speeds above average allows car drivers to reach
distant places over-proportionally fast, which results in higher
values of $\delta$. Thus, the effective dimension can be used as a
measure of the heterogeneity of the road speeds.
Fig.~\ref{eq:dimension}(a) shows the areas reachable from a central
node within different travel-time budgets.

\begin{figure}[t]
    \begin{center}
        \includegraphics[width=0.9\textwidth]{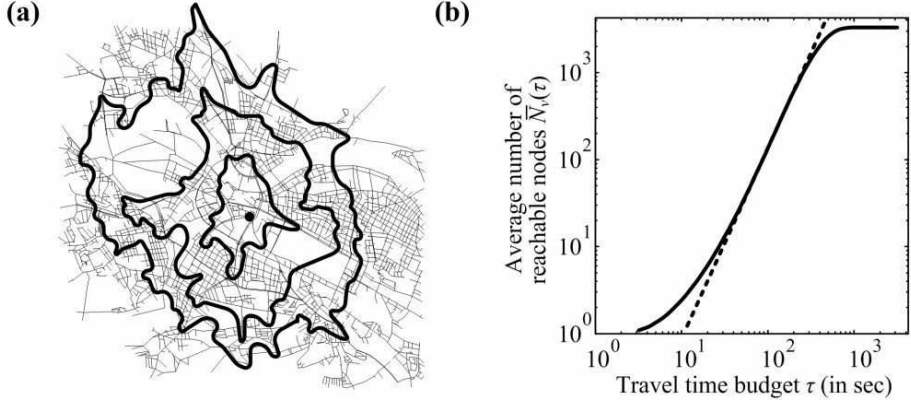}
    \end{center}
    \caption{
        (a) Isochrones (bold lines) surround areas reachable from
        a point in the city centre of Dresden with different
        travel-time budgets $\tau_1:\tau_2:\tau_3 = 1:2:3$.
        These areas extend wider along fast roads, e.g. in the north,
        while they are compressed along slower roads,
        e.g. in the east.
        (b) Average number $\overline{N}_v(\tau)$ of nodes
        reachable within a travel-time budget $\tau$.
    }
    \label{fig:effectiveDimension}
    \vspace{5mm}
\end{figure}

Referring to \cite{CsanyiSzendroi04,Gastner04}, the effective
dimension $\delta$ is theoretically defined by
\begin{equation}
    \label{eq:dimension}
    \delta = \lim_{\tau \rightarrow \infty}\,
    \frac{d \, \log N_v(\tau)}{d \, \log \tau} \;\;.
\end{equation}
Since road networks are finite, however, this formula has to be
approximated. For all nodes we have computed the average
$\overline{N}_v(\tau)$ and plotted it over $\tau$ in double
logarithmic scale as shown in Fig.~\ref{fig:effectiveDimension}(b).
For larger values of $\tau$, the curve saturates due to the finite
number of nodes in the graph. The slope of this curve at its
inflection point gives the lower bound for an estimation of $\delta$
(dotted line). Alternatively, one could also periodically continue
the graph, e.g. by mirroring a rectangular part of it and estimate
the limit for $\tau \rightarrow \infty$.

\newpage

\section{Distribution of Traffic}
\label{sec:traffic}

\begin{figure}[t]
    \begin{center}
        \includegraphics[width=\textwidth]{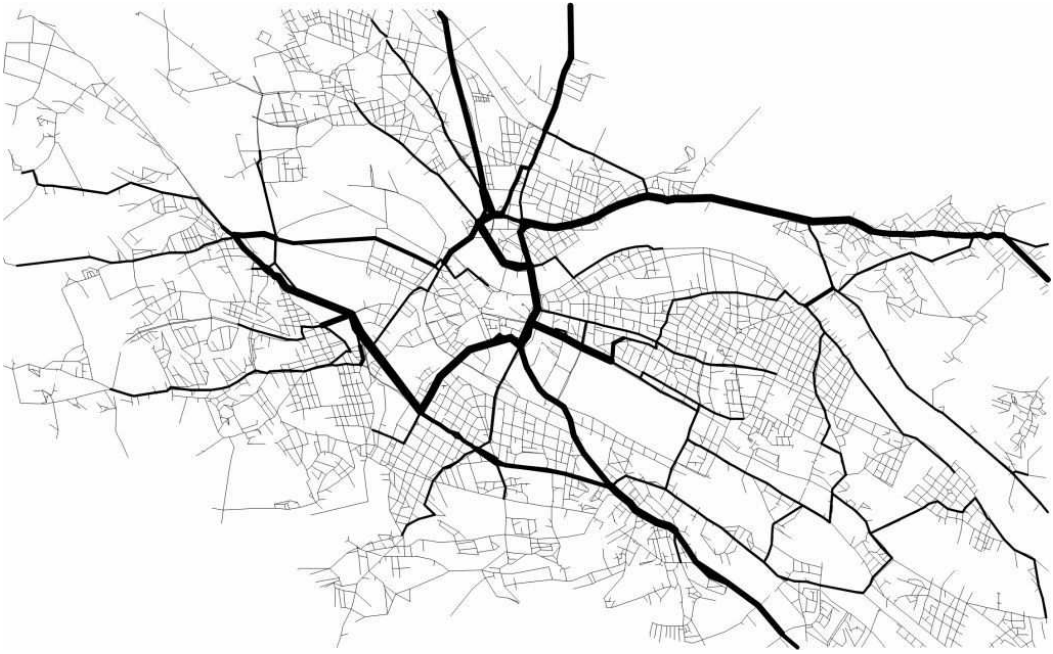}
    \end{center}
    \caption{
        Shortest paths in the road network of Dresden. The
        width of the links corresponds to the respective betweenness
        centrality $b_e$, that is an approximate measure of
        the amount of traffic on that roads.
    }
    \label{fig:traffic}
    \vspace{1cm}
\end{figure}

The heterogeneity of road speeds also has an impact on the
distribution of vehicular traffic in the road network. Faster roads
are more attractive for human drivers, resulting in a concentration
of traffic along these roads, see Fig.~\ref{fig:traffic}.

The importance of a road or a junction can be characterised by the
number of cars passing through it within some time interval. This
can roughly be approximated with the measure of link betweenness
centrality $b_e$ and node betweenness centrality $b_v$. It is given
by the number of shortest paths with respect to travel-time between
all pairs of nodes in the corresponding graph, of which the
particular link $e$ or node $v$ is part of
\cite{Albert02,Brandes05,Costa05,Newman02,PortaPrimal}. Using the
measure of betweenness centrality holds, we assume equally
distributed origin-destination pairs, identical average departure
rates, and the absence of externalities. Even though these
assumptions might not hold for precise traffic flow predictions,
they allow for estimating the implications of the network topology
on the spatial distribution of traffic flows.

\begin{figure}[t]
    \begin{center}
        \includegraphics[width=0.9\textwidth]{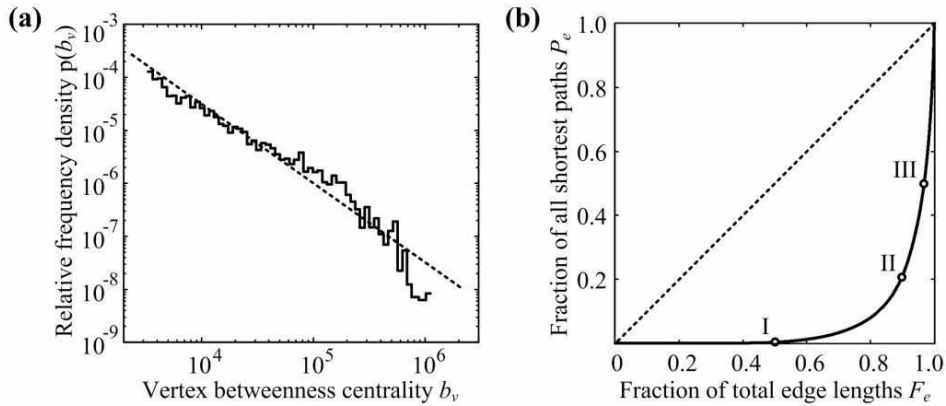}
    \end{center}
    \caption{
        (a) The distribution of the node betweenness centrality $b_v$
        obeys the power-law $p(b_v) \sim b_v^{-\beta}$
        with the exponent $\beta=1.355$ for Dresden (dotted line).
        (b) The Lorenz curve (solid line) for Dresden.
    }
    \label{fig:concentration}
    \vspace{1cm}
\end{figure}

The German road networks show an extremely high node betweenness
centrality $b_v$ at only a small number of nodes, while its values
are very low at the majority of nodes.
Fig.~\ref{fig:concentration}(a) shows the distribution of its
relative frequency density~$p(b_v)$. Over the entire range, the
distribution follows the scale-free power-law
$p(b_v)~\sim~b_v^{-\beta}$ with the exponent $\beta=1.355$ for
Dresden, see also Table \ref{tab:cities}. High values of $\beta$ can
be interpreted as a high concentration of traffic on the most
important intersections.

\newpage

Studying the link betweenness centrality $b_e$ reveals a similar
picture: The traffic volume is highly concentrated on only a few
roads, or to be more precise, on only a few road meters. By
referring to road meters instead of roads we overcome the effect of
different road lengths. As a quantitative concentration measure we
use the Gini index $g$, which can be obtained from the Lorenz curve
\cite{Lorenz1905}. The Lorenz curve is an monotonously increasing
and convex curve joining the points $(F, P)$, where $F$ is the
fraction of all road meters that have a fraction $P$ of the total
length of all shortest paths leading over it. The Gini index $g$ is
defined as twice the area between the Lorenz curve and the diagonal.
In the extreme case of a perfect equal distribution, the Lorenz
curve would follow the diagonal with $g=0$. In the other extreme
case of a distribution similar to delta function, we would find
$P=0$ for all $F<1$, and $P=1$ if $F$=1, and the Gini index would be
$g=1$. The Lorenz curve for the road network of Dresden is shown in
Fig.~\ref{fig:concentration}(b) and can be interpreted as follows:
50\% of all road meters carry as little as 0.2\% of the total
traffic volume only (I), while almost 80\% of the total traffic
volume are concentrated on no more than 10\% of the roads (II). Most
interestingly, half of the total traffic volume is handled by only
3.2\% of the roads in the network (III). The related Gini index of
Dresden is $g=0.870$, see Table \ref{tab:cities}.

\newpage

The bundling of traffic streams on a few arterial roads reflects the
clear hierarchical structure of the roads. The existence of
hierarchies is an inherent property of transportation networks
\cite{LevinsionYerra05}. Fig.~\ref{fig:traffic} shows that the
arterial roads sprawl out from the city centre in all direction of
the network.

Besides the diversity of road speeds, the inherent structure of the
road network topology itself has a tremendous effect on the
emergence of road hierarchies. Dead-end roads, for example, are at
the lowest level of the road hierarchy by definition. Interestingly,
the fraction of dead-end roads or, more precisely, the fraction of
tree-like structures in the corresponding graph, is about 20\% of
the total road length in the network of Dresden. Some of the
dead-ends may belong to the boundary of the road network, but their
fraction should be small since only a few country roads or highways
are cut. Such tree-like structures, also referred to as `lollipop'
layouts, are typical for modern North American suburbs
\cite{PortaPrimal} and are found among the 20 German cities under
consideration as well.

\section{Cellular Structures}
\label{sec:cells}

The structure and spatial extension of trail systems
\cite{Batty1997,HelbingKeltschMolnar97} is constrained by the
presence of impenetrable or inaccessible places. The structure of
road networks, therefore, is a result of an interplay between travel
cost minimisation and efficient land use. Facilities, residences,
parks etc. are enclosed by the roads, letting the road network
appear as a two-dimensional cellular system. Such structures are
typical for trail systems as well as for self-generated structures
like crack patterns, maple leaves, dragonfly wings etc.
\cite{Schaur92}.

\newpage

\begin{figure}[t]
    \begin{center}
        \includegraphics[width=0.9\textwidth]{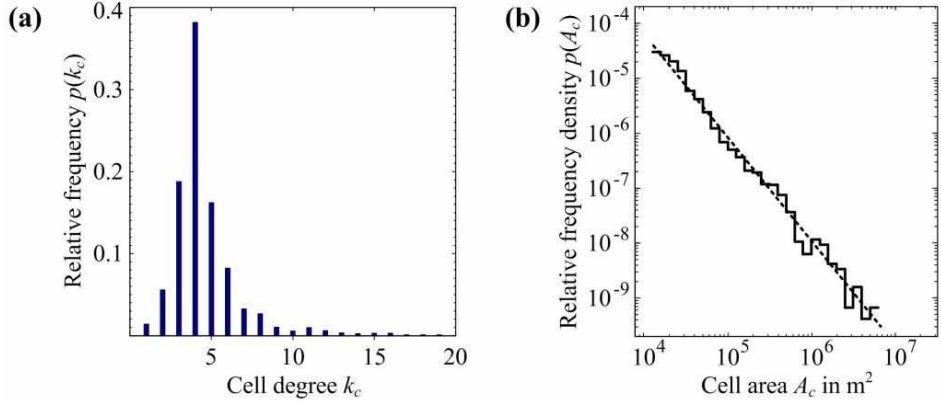}
    \end{center}
    \caption{
    (a) Frequency distribution of neighbourhood degrees $k_c$ of the cells
    in the road network of Dresden. The predomination of cells with
    four neighbours was found in all 20 German road networks.
    (b) The frequency distribution of the cell's surface areas $A_c$ obeys
    the scale-free power-law $P(A_c)~\sim~A_c^{-\alpha}$ (dotted line) with the
    exponent $\alpha=1.892$ for the road network of Dresden.
    }
    \label{fig:cells}
    \vspace{1cm}
\end{figure}

The topology of two-dimensional cellular structures has been studied
in the domain of planar graph theory \cite{Gibson99,Godreche92}
since Euler, whose theorem states that the number $N_c$ of bounded
cells in a connected planar graph with $N_v$ nodes and $N_e$
undirected links is given with $N_c = N_e - N_v + 1$. The graph of
road networks is always connected but, due to the presence of
bridges and tunnels, obviously not planar in a strict sense, as is
required for the definition of cells. Thus, for our investigations,
we determined all pairs of crossing links and connected them by
adding virtual nodes at the crossing points.

A cell's neighbourhood degree $k_c$ is the number of adjacent cells
\cite{Godreche92} or, which is equal to that, the number of
non-dead-end roads the cell is connected to. The frequency
distribution $P(k_c)$ of neighbourhood degrees for the road network
of Dresden is shown in Fig.~\ref{fig:cells}(a). In all 20 road
networks under consideration, around 80\% of the cells have three to
six neighbours, where those with four neighbours are always
predominating. This is in perfect agreement with the observations in
non-planned settlements, while in crack patterns and maple leaves
the most frequent neighbourhood degree is always five, in dragonfly
wings and honey combs it is even six \cite{Schaur92}. This leads to
the conjecture, that the most frequent neighbourhood degree of four
is a distinctive feature of urban road networks. The frequency
density distribution $p(A_c)$ of the surface areas $A_c$ is shown in
Fig.~\ref{fig:cells}(b). Note that we neglected cells of size
smaller than $10,000 \mathrm{m}^2$, which are usually artefacts of
the data's high precision, obviously representing vacancies within
more complicated intersection layouts. The distribution $p(A_c)$ is
scale invariant and obeys the power-law $p(A_c)~\sim~A_c^{-\alpha}$
with the exponent $\alpha=1.892$ for the road network of Dresden,
see Table~\ref{tab:cities}.

\newpage

\begin{figure}[t]
    \begin{center}
        \includegraphics[width=0.9\textwidth]{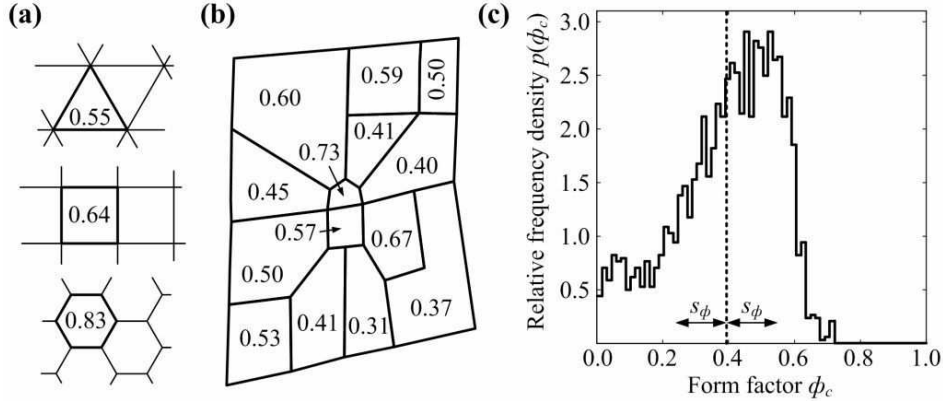}
    \end{center}
    \caption{
    Form factor $\phi_c$ of cells in (a) regular structures
    and (b) in the district of Striesen in the road network of Dresden.
    (c) The frequency density distribution $p(\phi_c)$ of the form factors
    has a standard deviation (small arrows) of $s_\phi=0.156$ indicating a broad
    diversity of the cell shapes. The dotted line represents the mean value $\bar{\phi}_c$.}
    \label{fig:formFactor}
    \vspace{1cm}
\end{figure}

As a quantitative measure of the compactness or roundness of a cell
$c$, we use the form factor $\phi_c$. It is the fraction of the
surface area of the circumscribed circle that is covered by the
cell. With $D_c$ denoting the maximum distance between two points on
the boundary of the cell, the form factor can be estimated by
$\phi_c = 4/\pi \left(A_c / D_c^2\right)$. The values of $\phi_c$
range from 0 to 1 and correspond to the limiting cases of infinitely
narrow and perfectly circular cells, respectively.
Fig.~\ref{fig:formFactor}(a) gives an example of form factors in
homogenous grid structures and Fig.~\ref{fig:formFactor}(b) shows a
small part of the road network of Dresden. The frequency density
distribution $p(\phi_c)$ of form factors in the road network of
Dresden is shown in Fig.~\ref{fig:cells}(c). The maximum value found
is $\phi_c=0.73$, while 70\% of the cells have a form factor in the
range between 0.3 and 0.6. The standard deviation of $s_\phi=0.156$,
see Table~\ref{tab:cities}, reflects a broad diversity of cell
shapes. This might result from the long history of German cities,
that were growing over several centuries and contain both, historic
centres and modern regularly structured areas designed according to
today's infrastructural demands.

\section{Summary}
\label{sec:summary}

We have analysed real-world data of urban road networks of the 20
largest German cities. Considering travel-times rather than
distances reveals an effective dimension significantly larger than
two. Centrality measures allow for the quantification of `important'
or frequently used road segments and reflect the hierarchical
structure of the road network. The shape of the cells encircled by
road segments can be quantified through the notion of a form factor.
We found scaling of several aspects of road networks, such as the
distribution of cell sizes or the number of nodes reachable within a
travel-time budget. In contrast to many material transport networks
in biology such as vascular \cite{BrownWest00}, however, their
topological organisation is less obvious and a hierarchical
structure similar to a Cayley tree is not found at all.

\section{Acknowledgement}
\label{sec:acknowledgement}

We thank Geoffrey West and Janusz Ho{\l}yst for inspiring
discussions, Winnie Pohl and Kristin Meier for their support of our
data analysis, and for partial financial support within the DFG
project He 2789/5-1. S.L. is grateful for a scholarship by the
`Studienstiftung des Deutschen Volkes'.




\end{document}